\documentclass[onecolumn,showpacs,preprintnumbers,amsmath,amssymb]{revtex4}
\usepackage[english]{babel}
\usepackage{graphics}
\usepackage{epsf}
\usepackage{psfig}
\usepackage{dcolumn}
\usepackage{bm}
\begin{document}

\title{An Optical Approach to the Dynamical Casimir Effect}

\author{Pawe\l{} W\c{e}grzyn}
\affiliation{ Marian Smoluchowski Institute of Physics,
Jagellonian University, Reymonta 4, 30-059 Cracow, Poland}
\email{wegrzyn@th.if.uj.edu.pl}

\begin{abstract}
We recently proposed a new approach to analyze the parametric
resonance in a vibrating cavity based on the analysis of classical
optical paths. This approach is used to examine various models of
cavities with moving walls. We prove that our method is useful to
extract easily basic physical outcome.
\end{abstract}


\maketitle

\section{Introduction}

The derivation of the Casimir effect in its original context
\cite{casimir} was the calculation of the electromagnetic force
between parallel perfectly conducting plates placed in a vacuum.
The geometry of a cavity formed by the parallel plates is also the
best context to manifest the nonstationary (dynamical) Casimir
effect. The theoretical calculations were first carried out by
Moore \cite{moore}. The force between plates is dynamically
modified, but one encounters more amazing phenomenon being the
vacuum emission of photons with nonclassical properties. The
latter topic became the subject of intensive studies since it was
recognized that the production of photons could be significantly
enhanced under the parametric resonance conditions \cite{law,
dodonov}. From the experimental point of  view, the hopes for
observing vacuum radiation are based on the belief that vibrating
cavities can be adjusted to parametric resonance conditions.

Quantum parametric resonance is a complex process and traditional
and well-tried methods for calculations in quantum electrodynamics
are obsolete here. We must examine open quantum systems under the
action of external periodic perturbations. Moreover, the focus of
our interest are the instable systems. If we describe physical
quantities in resonant systems, then the corresponding
mathematical  functions  inevitably yield narrow peaks or steep
stairs. A naive approach to make perturbations in small amplitudes
of cavity vibrations breaks down. Therefore, we must go beyond the
standard mathematical treatment known from QFT and pursue some new
machinery. Most ways of tackling the problem follow the
quantization scheme described by Moore \cite{moore} in case of
one-dimensional cavities. Then, special mathematical techniques
are used to trace the evolution of the quantum system with
time-dependent boundary conditions. Dodonov and his collaborators
\cite{review} reduced the analysis of field equations in a cavity
system to some infinite set of coupled and time-dependent
oscillators. Another idea based on a concept of renormalization
group was proposed in \cite{dalvit1, dalvit2}. Our suggestion
\cite{wegrzyn3} is to trace the evolution of the quantum fields in
a vibrating cavity system following classical optical paths. It is
related to the iteration procedure described in \cite{cole} and
techniques developed for classical field theory  in
\cite{dittrich} and \cite{meplan}. In fact, the resonant evolution
of the physical system is similar in classical and quantum models.
Recently, a similar idea to ground calculations  on the analysis
of classical optical paths was advocated for a stationary Casimir
effect  in \cite{jaffe1, jaffe2}.

In the present paper, we apply our technique \cite{wegrzyn3} to
examine cavities with various kinds of their motion and different
types of boundary conditions. The calculations are involved in
determining the essential physical outcome. In particular, the
conditions necessary to keep the parametric resonance  and the
critical exponents describing resonant evolution are derived. All
results are in agreement with those obtained for cavities with
mowing walls using other techniques. We give more information
about resonance windows, more subtle structure of energy densities
when amplitudes increase or the adjustment of resonance parameters
is getting worse and other new details.

\section{A cavity with one oscillating wall}

Most researches on vibrating cavities concentrated on a model of a
cavity formed by a static wall and a moving one. The static wall
is fixed at $x=0$, while the moving one follows its trajectory
$x=L(t)$. The condition that $L(t)>0$ is assumed. It is convenient
to make use of the billiard function \cite{wegrzyn1}:
\begin{equation}\label{bf}
f(t+L(t))=t-L(t) \ .
\end{equation}
This function contains all information about possible scatterings
of massless particles from the moving wall. It can be also
understood as the trajectory of the wall in light-cone
coordinates. Its derivatives are recognized as retarded Doppler
factors (the retardation relation is $t = t^\star+L(t^\star)$):
\begin{equation}\label{df}
    \dot{f}(t)=\frac{1-\dot{L}(t^\star)}{1+\dot{L}(t^\star)} \ .
\end{equation}
Consider now a classical scalar field $A(t,x)$ in a
one-dimensional cavity. It obeys the wave equation:
\begin{equation}
(\partial_t^2-\partial_x^2) A(t,x)=0 \ .
\end{equation}
The field is subject to some boundary conditions. Usually, we
assume either Dirichlet's boundary conditions:
\begin{equation}
A(t,0)=A(t,L(t))=0 ,
\end{equation}
or Neumann's boundary conditions at the instantaneous positions of
walls:
\begin{equation}
\partial_x A(t,0)=(\dot{L}(t) \partial_t - \partial_x) A(t,L(t))=0 ,
\end{equation}
In both cases, the classical solutions are given by:
\begin{equation}
A(t,x)=  \varphi(t+x) \mp \varphi(t-x),
\end{equation}
where the upper sign corresponds to Dirichlet's case, while the
lower one to Neumann's case. The profile function is subject to
the following relation:
\begin{equation}
\varphi(\tau)=\varphi(f(\tau)) \ .
\end{equation}
The energy density of the classical wave packet and its total
energy are given by:
\begin{equation}\label{ced}
T_{00}(t,x)=1/2(\partial_t A)^2+1/2(\partial_x
    A)^2=\varrho(t+x)+\varrho(t-x) \ ,
\end{equation}
\begin{equation}\label{te}
E(t)=\int_0^{L(t)} \ dx \ T_{00}(t,x)=\int_{t-L(t)}^{t+L(t)} \
d\tau \ \varrho(\tau)  \ ,
\end{equation}
where $\varrho(\tau)=\dot{\varphi}^2(\tau)$. Let us now describe
classical optical paths. We consider a massless particle that
starts at time $\tau$ moving right from the position $x=0$. Denote
by $T_n(\tau)$ and $T^\star_n(\tau)$ times of $n$-th collisions of
the particle with the moving wall and the static one respectively.
It is easy to notice that $T_n(\tau)= (f^{-1})^{\circ n}(\tau)$,
where the notation $(f^{-1})^{\circ n}(\tau)$ means here and
throughout this paper $n$-fold composition $f^{-1}\circ
f^{-1}\circ \ldots \circ f^{-1}$, while $T^\star_n(\tau)$ are to
be determined from the retardation relations
$T^\star_n(\tau)+L(T^\star_n(\tau))=T_n(\tau)$. The parametric
resonance is related to the existence of periodic particle
trajectories \cite{meplan, wegrzyn3}. Each periodic trajectory
obeys a following condition (for any non-negative integer $n$):
\begin{equation}\label{pt}
     T_n(\tau_0)=\tau_0+n T \ ,
\end{equation}
where $\tau_0$ is a starting point in time  and $T$ is a period.
The retarded times are respectively
$T^\star_n(\tau_0)=T_n(\tau_0)-T/2$.  Any periodic particle
trajectory appears on condition that the mirror trajectory  has
return points:
\begin{equation}\label{rp}
    L\left(\tau^\star_0+n T\right)=T/2 \ .
\end{equation}
It is usually assumed that  $T/2$ refers to the length of the
static cavity, so that $T/2\equiv L\equiv L(0)$. The crucial
characteristic of a particle trajectory is the cumulative Doppler
factor \cite{wegrzyn3}:
\begin{equation}\label{cdf}
D_n(\tau)= \frac{1}{\dot{T}_n(\tau)} =\prod_{k=1}^n
\dot{f}(T_n(\tau))= \prod_{k=1}^n
\frac{1-\dot{L}(T_k^\star(\tau))}{1+\dot{L}(T_k^\star(\tau))} \ .
\end{equation}
The above function has a direct physical interpretation. If a
massless particle starts at time $\tau$ moving right inside the
cavity, then after $n$ rebounds from both cavity walls its kinetic
energy changes by the factor $D_n(\tau)$.  A trajectory is called
positive (or negative respectively)  if its cumulative Doppler
factor tends to infinity (or zero) with increasing $n$. We denote
by $\tau_+$ and $\tau_-$ starting points for positive and negative
periodic trajectories. If  positive trajectories exist, then any
trajectory approaches one of the positive trajectories for long
times. The same is true for negative trajectories if we consider
the evolution backward in time. In other words, $T_n(\tau)
\rightarrow T_n(\tau_+)$ and  $T^{-1}_n(\tau) \rightarrow
T^{-1}_n(\tau_-)$ for large $n$.

The evolution of the energy density can be deduced from the
following formula,
\begin{equation}\label{crr}
\varrho(T_n(\tau))=\varrho(\tau)  D_n^2(\tau) \ ,
\end{equation}
together with the asymptotic approximation for long times (i.e.
$n\gg 1$)\cite{wegrzyn3}:
\begin{equation}\label{edas1}
\varrho(nT+\tau_++\varepsilon D_n^{-1}(\tau_+)) \cong
\varrho(\tau_++\varepsilon)D_n^2(\tau_++\varepsilon) \ ,
\end{equation}
\begin{equation}\label{edas2}
\varrho(nT+\tau_-+\varepsilon) \cong \varrho(\tau_-+\varepsilon
D_n(\tau_-))D_n^2(\tau_-+\varepsilon D_n(\tau_-)) \ .
\end{equation}
The above formulas explain the formation of travelling narrow
packets in the energy density $T_{00}(t,x)$
\cite{law,review,lambrecht} . The profile function $\varrho(\tau)$
has peaks  located at the spots of positive periodic trajectories
$\tau_+$.  The height of a peak grows like $D_n^2(\tau_+)$, while
its width diminishes like $D_n^{-1}(\tau_+)$. The value of the
energy density near the peak decreases like $D_n^2(\tau_-)$. The
evolution of the total energy can be analyzed using the formula
(we assume here $\tau_0 \leq L$):
\begin{equation}\label{cte}
E(T^\star_n(\tau_0))=\int_{T_{n-1}(\tau_0)}^{T_n(\tau_0)} \ d\tau
\ \varrho(\tau) = \int_{f(\tau_0)}^{\tau_0} \ d\tau \
\varrho(\tau) D_n(\tau) \ .
\end{equation}
To calculate the total accumulated energy inside the cavity after
some time we need  only an initial energy density (i.e.
$\varrho(\tau)$ for $\tau \leq L$) and the knowledge of cumulative
Doppler factors. In fact, it is usually enough to know these
factors in the vicinities of spots of periodic trajectories.

Now, we turn into the basic model of vibrating cavities specified
by the sinusoidal cavity wall motion:
\begin{equation}\label{sin}
    L(t)=L+\Delta L \sin{(\omega t)} \ ,
\end{equation}
where $\Delta L<L$ and $\omega\Delta L<1$. The parametric
resonance frequencies are $\omega_N=N\pi/L$, where $N$ is the
order of the resonance. Positive and negative periodic particle
trajectories correspond to the following starting points:
\begin{equation}
\tau_{+m}=\frac{(-N+2m+1)L}{N} \ \ \ , \ \ \
\tau_{-m}=\frac{(-N+2m)L}{N} \ ,
\end{equation}
where $m=0,1,...,N-1$. The cumulative Doppler factors can be
calculated:
\begin{eqnarray}
D_n(\tau_{\pm m})=\left(\frac{1\pm\omega_N\Delta
L}{1\mp\omega_N\Delta L}\right)^n  \nonumber \\
 \cong \exp{(\pm 2n\omega_N\Delta
L)} \quad {\rm for} \ \ \Delta L\ll L \ .
\end{eqnarray}
We conclude that there are $N$ travelling peaks in the energy
density. The average distances between peaks are equal.  For $N$
even, $N/2$ peaks are moving right and $N/2$ peaks are moving
left. For odd $N$ the number of peaks of each kind differs by one.
All the peaks have the same height and the same width. The total
energy grows exponentially with time. These classical results
agree with the results of the corresponding quantized cavity model
\cite{dodonov}. This fact will be explained later.

The above picture is true provided that the amplitude of cavity
oscillations is small. Strictly speaking, we need $\Delta L/L <
N$. For larger amplitudes, the behavior of the cavity system is
slightly modified. If $\Delta L/L
> M/N$ ($M=1,2,3,...$), then there appear additional positive
trajectories corresponding to the following return points
(\ref{rp}):
\begin{equation}
\omega_N \Delta L \sin{(\omega_N \tau^\star_{M\pm})}=M\pi \ ,
\end{equation}
and the corresponding cumulative Doppler factors are:
\begin{equation}
D_n(\tau_{M\pm})=\left(\frac{1\pm\sqrt{(\omega_N\Delta
L)^2-(M\pi)^2}}{1\mp\sqrt{(\omega_N\Delta L)^2-(M\pi)^2}}\right)^n
.
\end{equation}
It follows that beside the first series of $N$ peaks in the energy
density, there appear next $M$   series of $N$ peaks each. The
peaks of additional series are smaller and located between higher
peaks.

For practical purposes, it is important to consider off resonant
behavior of the cavity system. The instability in the field theory
appears not only for finely tuned frequencies. We are usually
dealing with some band structure. To establish the resonance
window, we move the resonant frequency: $\omega=\omega_N+\Delta
\omega$. The equation for return points (\ref{rp}) is now:
\begin{equation}
\sin(\omega \tau^\star)=-\frac{L\Delta\omega}{\omega\Delta L} \ ,
\end{equation}
so that  solutions  exist provided that:
\begin{equation}\label{bs}
    \frac{\Delta \omega}{\omega}<\frac{\Delta L}{L} \ .
\end{equation}
This condition agrees with the result of the  quantized model
\cite{wegrzyn2}. It defines the windows for the resonant
frequencies to keep the resonant instability. The cumulative
Doppler factors at starting points of periodic trajectories are:
\begin{equation}
D_n(\tau_{\pm m})=\left(\frac{1\pm\sqrt{(\omega\Delta
L)^2-(L\Delta\omega)^2}}{1\mp\sqrt{(\omega\Delta
L)^2-(L\Delta\omega)^2}}\right)^n .
\end{equation}
Again, the above picture is exact provided that $\Delta
L/L+(1+1/N)\Delta\omega/\omega>1/N$. Otherwise, next series of
smaller peaks in the energy density will appear.

Now, we present quantum formulas. The vacuum expectation value of
the energy density is given by \cite{moore}:
\begin{equation}
\langle T_{00}(t,x)\rangle = \varrho(t+x)+\varrho(t-x) \ ,
\end{equation}
where
\begin{equation}
\varrho(\tau)=-\frac{\pi}{48}
\dot{R}^2(\tau)-\frac{1}{24\pi}S[R](\tau) \ ,
\end{equation}
and the Schwartz derivative is defined by:
\begin{equation}
S[R]=\frac{\stackrel{...}{R}}{\dot{R}}-\frac{3}{2}
\left(\frac{\stackrel{..}{R}}{\dot{R}}\right)^2 \ .
\end{equation}
The function $R(\tau)$ is subject to Moore's equation:
\begin{equation}\label{me}
    R(\tau)-R(f(\tau))=2 \ ,
\end{equation}
where our billiard function appears in the quantum case as well.
The formula for the evolution of the energy density is given by:
\begin{equation}\label{rr}
\varrho(T_n(\tau))=\varrho(\tau)D^2_n(\tau)+A_n(\tau)D^2_n(\tau) \
,
\end{equation}
where the cumulative Doppler factor is defined again by the same
formula Eq.~(\ref{cdf}) and the definition of the quantum term
called the cumulative conformal anomaly contribution
\cite{wegrzyn3} is  given below:
\begin{equation}\label{cac}
    A_n(\tau)= \frac{1}{24\pi} S[T_n](\tau)=
     - \frac{1}{24\pi}\sum_{k=1}^n
    D_k^{-2}(\tau)S[f](T_k(\tau)) \ .
\end{equation}
The quantum formula for the total energy accumulated in the cavity
is:
\begin{eqnarray}\label{qte}
E(T^\star_n(\tau_0))=\int_{T_{n-1}(\tau_0)}^{T_n(\tau_0)} \ d\tau
\ \varrho(\tau) \nonumber \\
 = \int_{f(\tau_0)}^{\tau_0} \ d\tau \ \left[
\varrho(\tau)+ A_n(\tau) \right] D_n(\tau) \ .
\end{eqnarray}
The quantum anomaly implies the difference between classical and
quantum formulas for the evolution of the energy. However, under
the resonance conditions the leading role is played by the
cumulative Doppler factor. This fact explains why  most of
classical results match quantum ones. It refers to the band
structure of resonance frequencies, the formation and the shape of
travelling packets in the energy density, the exponential growth
of the total energy. To make predictions for these features,  the
classical theory could be good enough. But it is not true for the
lowest resonance channel. In this case,  the anomalous mechanism
of energy growth clashes with the resonant enhancement of the
initial vacuum fluctuations \cite{wegrzyn3}. The lack of the
instability for the lowest resonance frequency is also related to
the $SL(2,R)$ symmetry of the cavity system \cite{wu2,wegrzyn4},
which is the remainder of the conformal group.

Finally, let us also conclude from the above results that there is
no difference between imposing Dirichlet's and Neumann's boundary
conditions in the context of our cavity models.

\section{A cavity with two oscillating walls}

A natural generalization of the considerations of the previous
section is to allow both walls to oscillate \cite{review, dalvit2,
ji}. Consider now that the right cavity wall moves along the
trajectory $x=L_1(t)$, while the left one follows $x=-L_2(t)$. We
assume also that the cavity was static in the past, so that
$L_1(t)=L_2(t)=L/2$ for $t<0$, and $L$ corresponds to the static
cavity length. The general formulas given in this section will be
rigorously justified if we assure that $L_1(t)>0$ and $L_2(t)>0$
(the walls never collides) and  the wall velocities are never
close to the velocity of light. The billiard functions are
implemented for both trajectories ($i=1,2$):
\begin{equation}\label{bf2}
f_i(t+L_i(t))=t-L_i(t) \ .
\end{equation}
Moreover, it is convenient to define an additional pair of
billiard functions:
\begin{equation}\label{bf22}
f_L=f_2 \circ f_1 \ \ , \ \ \ \ f_R=f_1 \circ f_2 \ .
\end{equation}
They describe double reflections of massless particles from the
cavity walls. The set of classical optical paths is now more
complex. Consider again a massless particle that starts at time
$\tau$ moving from the position $x=0$. Denote by $T_{Ln}(\tau)$
(and $T_{Rn}(\tau)$ respectively) times when the particle returns
to its initial position $x=0$ after  $n$ rebounds from both walls
assuming that the first collision is with the left (or with the
right) wall. It is easy to find that $T_{L(R)n}(\tau)=
(f_{L(R)}^{-1})^{\circ n}(\tau)$. Further, $T^\star_{L(R)n}(\tau)$
and $T^{\star\star}_{L(R)n}(\tau)$ are times of successive
collisions with the left (right) and the right (left) walls for
both types of trajectories respectively. They can be derived from
the following relations:
\begin{equation}
T^\star_{L(R)n}(\tau)+L_{1(2)}(T^\star_{L(R)n}(\tau))
=T_{L(R)n}(\tau) \ ,
\end{equation}
\begin{equation}
T^{\star\star}_{L(R)n}(\tau)+L_{2(1)}(T^{\star\star}_{L(R)n}(\tau))
=f_{1(2)}(T_{L(R)n}(\tau)) \ .
\end{equation}
For both types of trajectories cumulative Doppler factors are
defined:
\begin{eqnarray}
D_{L(R)n}(\tau)= \frac{1}{\dot{T}_{L(R)n}(\tau)} =\prod_{k=1}^n
\dot{f}_{L(R)}(T_{L(R)}(\tau)) \nonumber \\
 = \prod_{k=1}^n
\frac{1-\dot{L}_{1(2)}(T_{L(R)k}^\star(\tau))}
{1+\dot{L}_{1(2)}(T_{L(R)k}^\star(\tau))} \prod_{k=1}^n
\frac{1-\dot{L}_{2(1)}(T_{L(R)k}^{\star\star}(\tau))}
{1+\dot{L}_{2(1)}(T_{L(R)k}^{\star\star}(\tau))}
 \ .
\end{eqnarray}
Classical solutions of a wave equation inside a cavity with two
moving walls are given by:
\begin{equation}
A(t,x)=  \varphi_L(t+x) \mp \varphi_R(t-x),
\end{equation}
where again the upper sign corresponds to Dirichlet's boundary
conditions and the lower one to Neumann's boundary conditions on
the walls. The profile functions in both cases obey the same
relations:
\begin{equation}
\varphi_L(\tau)=\varphi_R(f_1(\tau)) \ , \ \ \ \
\varphi_R(\tau)=\varphi_L(f_2(\tau)) \ .
\end{equation}
The energy density of the classical wave packet and its total
energy can be represented as:
\begin{equation}
T_{00}(t,x)= \varrho_L(t+x)+\varrho_R(t-x) \ , \ \ \
\varrho_{L(R)}(\tau)=\dot{\varphi}_{L(R)}^2(\tau) \ ,
\end{equation}
\begin{equation}
E(t)=\int_{-L_2(t)}^{L_1(t)} \ dx \
T_{00}(t,x)=\int_{t-L_2(t)}^{t+L_1(t)} \ d\tau \ \varrho_L(\tau) +
\int_{t-L_1(t)}^{t+L_2(t)} \ d\tau \ \varrho_R(\tau) \ .
\end{equation}
The classical evolution of the system can be traced with the help
of the following formulas:
\begin{equation}
\varrho_{L(R)}(T_{L(R)n}(\tau))= \varrho_{L(R)}(\tau)
D_{L(R)n}^2(\tau) \ ,
\end{equation}
\begin{equation}
E(t)=\int_{f_2(\tau_2)}^{\tau_1} \ d\tau \ \varrho_L(\tau)
D_{Ln}(\tau) + \int_{f_1(\tau_1)}^{\tau_2} \ d\tau \
\varrho_R(\tau) D_{Rn}(\tau)  \ ,
\end{equation}
where $\tau_{1(2)} \equiv f^{\circ n}_{L(R)}(t+L_{1(2)}(t))$. If
we take $n$ big enough to support that $\tau_{1(2)} \leq L/2$,
then we obtain the formula that allows us to calculate the
evolution of the total energy from the knowledge of Doppler
factors and the energy distribution in the initial state.

Before we turn into examples of cavity systems, we explain why
many specific systems of cavities with two oscillating walls just
reproduce results similar to those obtained in cavity models with
one oscillating wall. The reason is that billiard functions
(\ref{bf22}) correspond to some physical trajectories in two
dimensions. It means that we can find a trajectory $x=L(t)$ which
describes some realistic motion of the wall such that:
\begin{equation}
f_{L(R)}(t+L(t))=t-L(t) \ .
\end{equation}
We can verify that $L(t)>0$, $L(t)=L$ for $t<0$ and $\dot{L}(t)$
is never close to the speed of light. Obviously, in general such
"effective" trajectories are different for $f_L$ and $f_R$, but in
both cases the following composition rules are satisfied:
\begin{equation}\label{cr1}
L(t)=L_1(t_1)+L_2(t_2) \ ,
\end{equation}
\begin{equation}\label{cr2}
\frac{1-\dot{L}(t)}{1+\dot{L}(t)}
=\frac{1-\dot{L}_1(t_1)}{1+\dot{L}_1(t_1)}
\frac{1-\dot{L}_2(t_2)}{1+\dot{L}_2(t_2)} \ ,
\end{equation}
where $t\pm L(t)=t_1\pm L_1(t_1)$ and $t_1\mp L_1(t_1)=t_2\pm
L_2(t_2)$ for $f_L$ (the upper sign) and $f_R$ respectively.

The latter relation Eq.(\ref{cr2}) states just that $\dot{L}(t)$
is the relativistic sum of velocities $\dot{L}_1(t_1)$ and
$\dot{L}_2(t_2)$. However, this composition rule has nothing to do
with any relative motion since there exists no inertial frame
where we could make respective time moments synchronous.
Nevertheless, both left- and right-movers in cavities with two
oscillating walls correspond to some models of cavities with a
single oscillating wall. In other words, each mode of a cavity
system with two moving walls can be described by a simpler model
discussed in the previous section.

Let us discuss   specific types of symmetric and antisymmetric
oscillations of cavities. In the paper of Dalvit and Mazzitelli
\cite{dalvit2}, these types of cavity oscillations are called
breathing modes ("electromagnetic antishaker") and translational
modes ("electromagnetic shaker"). For symmetric oscillations we
have $L_1(t)=L_2(t)$ and the cavity oscillates symmetrically with
respect to the center of the cavity. The billiard functions and
Doppler factors are identical, namely $f_1 \equiv f_2$, $f_L
\equiv f_R$ and $D_L \equiv D_R$. The model is equivalent to the
cavity model with one oscillating wall, where the prescribed
trajectory of the wall is defined by the billiard function $f_1
\circ f_1$. For antisymmetric oscillations we have
$L_1(t)+L_2(t)=L$, and the cavity is oscillating as a whole with
its length kept constant in the laboratory frame. Then,
$f_{1(2)}(\tau)=f_{2(1)}^{-1}(\tau-L)-L$ and
$f_{L(R)}(\tau)=f_{1(2)}^{-1}(f_{1(2)}(\tau)-L)-L$. It is
interesting to note that for periodic translational cavity
oscillations with $L$ being a period, the billiard functions $f_L$
and $f_R$ correspond to static ones. It follows that there is no
effect of the cavity movement on the energy distribution of fields
and the whole cavity system is equivalent to the static one.

Finally, we analyze harmonic oscillations of cavity walls:
\begin{eqnarray}\label{dho}
L_1(t)=\frac{L}{2}+\Delta L_1 \sin{(\omega_R t)} \ , \nonumber \\
L_2(t)=\frac{L}{2}+\Delta L_2 \sin{(\omega_L t-\delta)}+\Delta L_2
\sin{\delta} \ .
\end{eqnarray}
Again, we want to establish resonance windows, find the evolution
of the total energy and describe the shape of the local energy
density. We must first detect positive and negative periodic
trajectories of massless particles moving inside the cavity. They
can be derived from the generalization of Eq.(\ref{rp}):
\begin{equation}\label{rp2}
L_1(t_1)+L_2(t_1\pm \frac{T}{2}) = \frac{T}{2} \ ,
\end{equation}
where the upper (lower) sign corresponds to the "left" $T_{Ln}$
("right" $T_{Rn}$) trajectories. The cumulative Doppler factors at
the starting points of periodic trajectories can be calculated
then:
\begin{eqnarray}
D_{L(R)n}=\left(\frac{1-\omega_R \Delta L_1 \cos{(\omega_R
t_1)}}{1+\omega_R \Delta L_1 \cos{(\omega_R t_1)}} \right)^n
\left( \frac{1-\omega_L \Delta L_2 \cos{(\omega_L t_1 \mp \omega_L
T/2-\delta)}}{1+\omega_L \Delta L_2 \cos{(\omega_L t_1 \mp
\omega_L T/2-\delta)}} \right)^n \nonumber \\
\cong \exp{\left( -2n [\omega_R \Delta L_1 \cos{(\omega_R t_1)}
+\omega_L \Delta L_2 \cos{(\omega_L t_1 \mp \omega_L T/2-\delta)}]
\right)} \ \ \ {\rm for} \ \ \Delta L_i \ll L \ .
\end{eqnarray}
First, consider resonant oscillations with
$\omega_L=\omega_R=\omega_N\equiv N\pi/L$. If there is no
dephasing $\delta=0$, the system is equivalent to the cavity with
one oscillating wall. The oscillation frequency is $\omega_N$, the
effective cavity length is $L$ and the effective amplitude of
oscillations is $\Delta L_1+(-1)^N \Delta L_2$. For $\Delta
L_1=\Delta L_2$ and $N$ odd the effective cavity system is static.
The same is true for $\Delta L_1=\Delta L_2$, $\delta=\pi$ and $N$
even. These results agree with the paper \cite{dalvit2}. If the
amplitudes of oscillations are equal $\Delta L_1=\Delta L_2 \equiv
\Delta L$ and $\delta\neq 0, \pi$, then the solutions of
Eq.(\ref{rp2}) are $t_1=(\delta L/\pi+2 k)L/N$ and
$t_1=(N-2k-1)L/N$, where $k=0,1,...,N-1$. The respective
cumulative Doppler factors are:
\begin{equation}
D_n(\tau_{k\pm}) \cong \exp{\left( \pm 2n\omega_N \Delta
L(1+(-1)^N \cos{\delta}) \right)} \ .
\end{equation}
There are $N$ positive and $N$ negative particle trajectories. It
follows that there are $N$ peaks in the energy density. The
calculations for $\Delta L_1 \neq \Delta L_2$ are a bit more
complicated. However, if $\Delta L_1+(-1)^N \Delta L_2
\cos{\delta} \neq 0$ then there exist $N$ positive and $N$
negative particle trajectories. The corresponding cumulative
Doppler factors are given by:
\begin{equation}
D_n \cong \exp{\left( \pm 2n\omega_N (\Delta L_1+(-1)^N \Delta L_2
\cos{\delta}) \right)} \ .
\end{equation}
For $\Delta L_1+(-1)^N \Delta L_2 \cos{\delta} = 0$ there is no
exponential instability in the cavity system.

To establish the resonance window we assume
$\omega_L=\omega_R\equiv \omega_N+\Delta \omega$. The relation
Eq.(\ref{rp2}) yields:
\begin{equation}
\Delta L_1 \sin{(\omega t_1)}+(-1)^N \Delta L_2 \sin{(\omega
t_1-\delta)}+\Delta L_2 \sin{\delta}=\frac{-L\Delta
\omega}{\omega} \ .
\end{equation}
To keep the parametric resonance conditions for the vibrating
cavity Eq.(\ref{dho}) it is enough to support that:
\begin{equation}
\frac{\Delta\omega}{\omega}<\frac{\Delta L_2
\sin{\delta}+\sqrt{(\Delta L_1)^2+(\Delta L_2)^2+2(-1)^N\Delta
L_1\Delta L_2\cos{\delta}}}{L} \ .
\end{equation}

The quantum formulas for cavities with two oscillating walls
include anomaly contributions in the respective way. Since the
modifications are straightforward, we skip their discussion here.

\section{Conclusions}

An optical approach to the dynamical Casimir effect based on the
construction of classical optical paths is useful. In particular,
it is a powerful technique to analyze resonance systems. We can
trace the behavior of a cavity system with moving walls when we
increase amplitudes or loose fine tuning for resonance
eigenfrequencies. We can calculate modified critical exponents,
find more subtle structure of travelling peaks in energy densities
and control the conditions for the instability of the system. We
have established some "composition laws" coming into effect if we
want a more complicated model of a cavity with two moving walls
reduce to the analysis of a cavity with one moving wall. Moreover,
our understanding of the mechanism of parametric resonance in
cavities with moving walls and the implications of quantum field
theory is much better. For example, we can judge when some
semi-classical approximation is accurate or we can explain why
some resonances are not observed at the quantum level. It is
further remarkable that we can distinguish between exponential and
power-like instability and observe the role of symmetries at the
quantum level \cite{wegrzyn4}.

 We have seen in several examples that many qualitative or even
quantitative conclusions can be drawn from classical versions of
respective cavity models. Presumably, the parametric resonance
mechanism and evolution laws are similar in both cases. However,
we should be careful about quantum anomalies that can
significantly modify our predictions. A subtle role of quantum
modifications can be traced within our approach.

The results derived for Dirichlet's and Neumann's boundary
conditions for one-dimensional cavities are the same. This point
was already made in \cite{alves}. The cavities with two
oscillating walls possess a double set of characteristic
functions, namely two families of both cumulative Doppler factors
and cumulative anomaly contributions. It is related to the fact
that from a given space-time point there are two independent
directions to follow trajectories of massless particles. In three
dimensions, such functions depend also on localizations and
orientations.


\end{document}